\documentclass{article}

\usepackage{arxiv}

\usepackage[utf8]{inputenc} 
\usepackage[T1]{fontenc}    
\usepackage{hyperref}       
\usepackage{url}            
\usepackage{booktabs}       
\usepackage{amsfonts}       
\usepackage{nicefrac}       
\usepackage{microtype}      
\usepackage{lipsum}
\usepackage{graphicx}
\graphicspath{ {./images/} }

\title{Predicting Surface Heat Flux on Complex Systems via Conv-LSTM}

\author{
 Yinpeng Wang \\
  School of Electronic and Information\\
  Beihang University\\
  Beijing, 100191 \\
  \texttt{16081072@buaa.edu.cn} \\
   \And
 Nianru Wang \\
  School of Electronic and Information\\
  Beihang University\\
  Beijing, 100191 \\
  \texttt{17373024@buaa.edu.cn} \\
  \And
 Qiang Ren \\
  School of Electronic and Information\\
  Beihang University\\
  Beijing, 100191 \\
  \texttt{qiangren@buaa.edu.cn} \\
}

\begin{document}
\maketitle
\begin{abstract}
Existing algorithms with iterations as the principle for 3D inverse heat conduction problems (IHCPs) are usually time-consuming. With the recent advancements in deep learning techniques, it is possible to apply the neural network to compute IHCPs. In this paper, a new framework based on Convolutional-LSTM is introduced to predict the transient heat flux via measured temperature. The inverse heat conduction models concerned in this work have 3D complex structures with non-linear boundary conditions and thermophysical parameters. In order to reach high precision, a forward solver based on the finite element method is utilized to generate sufficient data for training. The fully trained framework can provide accurate predictions efficiently once the measured temperature and models are acquired. It is believed that the proposed framework offers a new pattern for real-time heat flux inversion.
\end{abstract}


\section{Introduction}
Real-time identifications of the heat flux distribution at a given domain are essential in many fields like engine design, geological prospecting and so on. Direct ways of heat flux measurements to the external surface of a hypersonic aircraft are expensive and difficult to maintain a high accuracy due to the severe condition \cite{najafi2015online}. As a result, there is a strong demand to develop a more precise and more efficient measurement technique for determining the heat flux distribution mediately. One reasonable approach is to take advantage of the numerous temperature measurement on the interior surface of the capsule to invert the heat flux on the exterior surface. Measuring the temperature is more economical and robust than directly measuring the heat flux. Hence, we need to solve the inverse heat conduction problems (IHCPs) to acquire the anticipated heat flux. 

An IHCP generally refers to the problems in identifying the unknown boundary or the thermal properties in a thermal system. As this problem is super sensitive to the measurement variation, it is mathematically ill-posed \cite{beck1985inverse}. Therefore, many valuable techniques have been proposed to transform it into a relatively well-posed problem. Traditional algorithms used in the field of IHCPs include sequential function specification method (SFSM) \cite{chantasiriwan1999comparison}, iterative techniques \cite{alifanov1974solution,bergagio2018iterative,cui2016modified,duda2016general,beck1982efficient}, Tikhonov regularization \cite{tikhonov1977solutions,huang2018estimation}, singular value decomposition with model-reduction \cite{shenefelt2002solution}, digital filter method \cite{najafi2015online,chen2017nonlinear,kowsary2006training}, genetic algorithm \cite{czel2012inverse,woodbury2000neural}, Bayesian framework \cite{zeng2019novel}, neural networks \cite{soeiro2004using, krejsa1999assessment, ghadimi2015heat, cortes2007artificial, mirsepahi2013comparative, czel2014simultaneous, czel2013inverse} and so on.

Many of the inverse techniques mentioned above perform well in heat conduct systems with simple geometries (1D or 2D models) and linear thermal parameters. However, in many practical cases, the problems are usually more sophisticated. Some works about those IHCPs with nonlinearity have been published. A regular way is to divide the nonlinear model into several linearized sub models. Keanini \textit{et al.} \cite{keanini2005modified} proposed a method based on the non-iterative finite element algorithms and successfully reconstruct the time-varying heat flux. The sequential function specification algorithm was adopted in the study of Huang \textit{et al.} \cite{huang2018estimation} predict the spatial and temporal related heat flux distribution by supposing that changes are linear at every time step while the sensitivity coefficients were obtained by finite element method calculations. Lv \textit{et al.} \cite{lv2019estimation} proposed a multiple model adaptive inverse method to forecast the heat flux condition of a nonlinear IHCP. In fact, the accuracy of sub-model algorithms relies on how many sub-models are constructed. It is hard for these techniques to solve IHCPs with severe nonlinearity, as they should use a considerable amount of sub-models and may cost unaffordable computational resources. A novel approximate Bayesian computation algorithm is proposed by Zeng \textit{et al.} \cite{zeng2019novel} for both linear and non-linear systems, while massive samples are needed. Therefore, it is hard for this method to solve problems within seconds. Duda \textit{et al.} \cite{duda2019new, duda2018solution}. efficiently solved the nonlinear IHCPs by dividing a complex domain into sample parts. However, the cases are limited to 2D models and the number of sensors needs to be specified in different cases. 

In industrial applications, the nonlinearity of the IHCP are more severe due to the following reasons: (1) The geometry of the model can be three-dimensional with intricate shapes and curved surfaces. In most practical cases, it is not appropriate to simplify the model into 1D or 2D models as it will lead to unbearable errors. (2) The nonlinearity of the model should be considered, especially the sophisticated radiation boundary condition and temperature-dependent materials. (3) The heat flux computation should be efficient, especially in a long-period inversion process. In such a case, the iteration-based methods with finite element methods solving forward problem \cite{bergagio2018iterative, duda2016general} are improper because of their unbearable time cost. In summary, handling IHCPs involving all the above aspects are much more meaningful but more complicated. Therefore, an efficient inverse algorithm is proposed to solve IHCPs with those characteristics.

The neural network has two advantages compared to the traditional method. One is that a fully trained framework spends less time to forecast the heat flux. The other is that the neural network can simulate the nonlinearity of the IHCP better than the traditional methods. Most of the studies have applied the artificial neural network (ANN) to solve IHCPs and used a backpropagation algorithm to optimize the framework. Ghadimi \textit{et al.} \cite{ghadimi2015heat} combined the Sequential Function Specification method and ANNs to handle the heat flux inversion of a locomotive brake disc. Via two layers of the fully connected network, he succeeded to forecast the linearized heat flux which is homogeneously distributed. The temporal-related heat flux is also predicted via perturbed temperature by Cortés \textit{et al.} \cite{cortes2007artificial}. They also used the fully connected network to capture the temporal features of the IHCP and simplified the question by assuming that the heat flux is homogenously located. However, in real industrial scenarios, it is impractical to utilize fully connected networks to solve complex IHCPs since the trainable parameters are too large. Thanks to the development of the research on deep learning, some special neural networks have been proposed.  The long short-term memory networks (LSTM) proposed by Hochreite \textit{et al.} \cite{hochreiter1997long} is an efficient framework handling these temporal-related systems. LSTM has been proven to be stable and powerful for modeling long-range dependencies in various precious studies as a special RNN structure \cite{graves2013generating, pascanu2013difficulty, sutskever2014sequence}. One superiority of employing the memory cell and gates to bridle information flow is that the gradient will be saved in the cell and be protected from fading too quickly while this is an essential problem for the vanilla RNN model \cite{graves2013generating, pascanu2013difficulty}. Multiple LSTMs can be coded and temporally linked to propose more complex structures, which have already been applied to solve many real-time sequence modeling problems \cite{xu2015show}. However, such a network can only accept 1D tensors, so it could only solve problems that contain information about time. Very recently, a new method is proposed by Shi \textit{et al.} \cite{shi2015convolutional} for precipitation nowcasting which modifies the common LSTM and puts the convolution into the layer to change the input tensor and adds the spatial information. Hence, the convolutional LSTM (ConvLSTM) is introduced as the base of our framework.

This paper is organized as follows. The second section firstly introduces the physical model of the IHCP to be investigated. Then the workflow and the method to generate the data set are illustrated. Furthermore, the architecture of the framework is also exhibited in detail. The third section displays and analyzes the outcome of the proposed framework. A brief conclusion is provided in the last section.

\begin{table*}[htbp]
\caption{Nomenclature}
\centering
\begin{tabular}{|cccc|}\hline

\textbf{Heat conduction}    &  & $\left[ {{K_1}^{(e)}} \right]$ & Coefficient matrix                \\
$\Omega$    & Solid body domain     & ${\left\{\overline P \right\}}$    & Load matrix    \\
$r$       & Space coordinate   & \textbf{ConvLSTM}    &  \\
${{S}_{1,2,3}}$    & Boundary surface   & ${{\chi }_{1}},\cdots ,{{\chi }_{t}}$    & Inputs of the ConvLSTM \\
${{T}_{0}}$       & Initial temperature   & ${{C}_{1}},\cdots ,{{C}_{t}}$     & Cell outputs of the ConvLSTM        \\
$t$    & Time     & ${{H}_{1}},\cdots ,{{H}_{t}}$     & Hidden states of the ConvLSTM                 \\
$q$     & Heat flux   & ${{i}_{t}},{{f}_{t}},{{o}_{t}}$  & Inside gates of the ConvLSTM         \\
${{T}_{c}}$   & Measured temperature    & $W$    &Coefficient of the ConvLSTM    \\
$k(T)$     & Thermal conductivity   & \textbf{Subscript}     &    \\
${{C}_{p}}(T)$   & Heat capacity     & $j$     & Sensor index    \\
$\rho $     & Mass density   & $w$  & Temperature on the radiation boundary     \\
$\sigma $   & Stefan Boltzmann constant   & $x$    & Input index \\
$\varepsilon $     & Emissivity  & $i,f,o$     & Gates index      \\
\textbf{FEM} & &$H$& Hidden states index \\
$N_i$ & Weight of the $i_{th}$ node & $C$  & Cell output index \\

\hline
\end{tabular}
\label{Table1}
\end{table*}

\section{Methodology and Formulation}
\begin{figure*}[htbp]
\centering
\includegraphics[scale=.45]{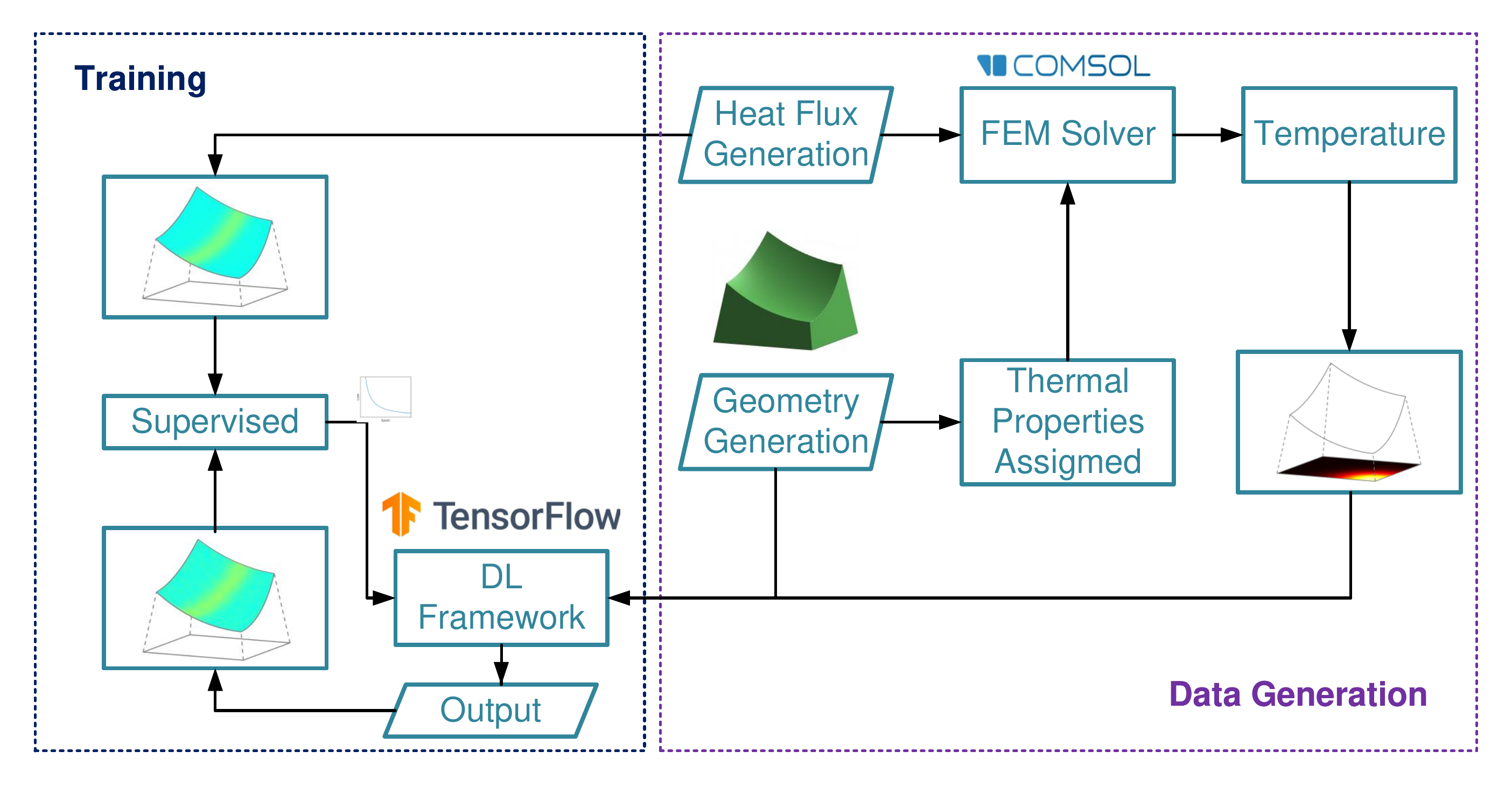}
\caption{The workflow to train and test the framework.}
\label{Fig1}
\end{figure*}

The whole workflow of the inversion process is demonstrated in Fig. \ref{Fig1}. Here, we first generate various geometries in MATLAB and combine the thermal properties with the geometries to produce the physical models. Then, the FEM solver utilizes the physical models with the heat flux generated by MATLAB to calculate the temperature distribution. After that, the generated geometries and temperatures in each time step are packed together to serve as the input of the framework. The framework is then optimized in the training process where the output is supervised by the generated heat flux.

\subsection{Nonlinear heat conduction model}
\begin{figure}[htbp]
\centering
\includegraphics[scale=.7]{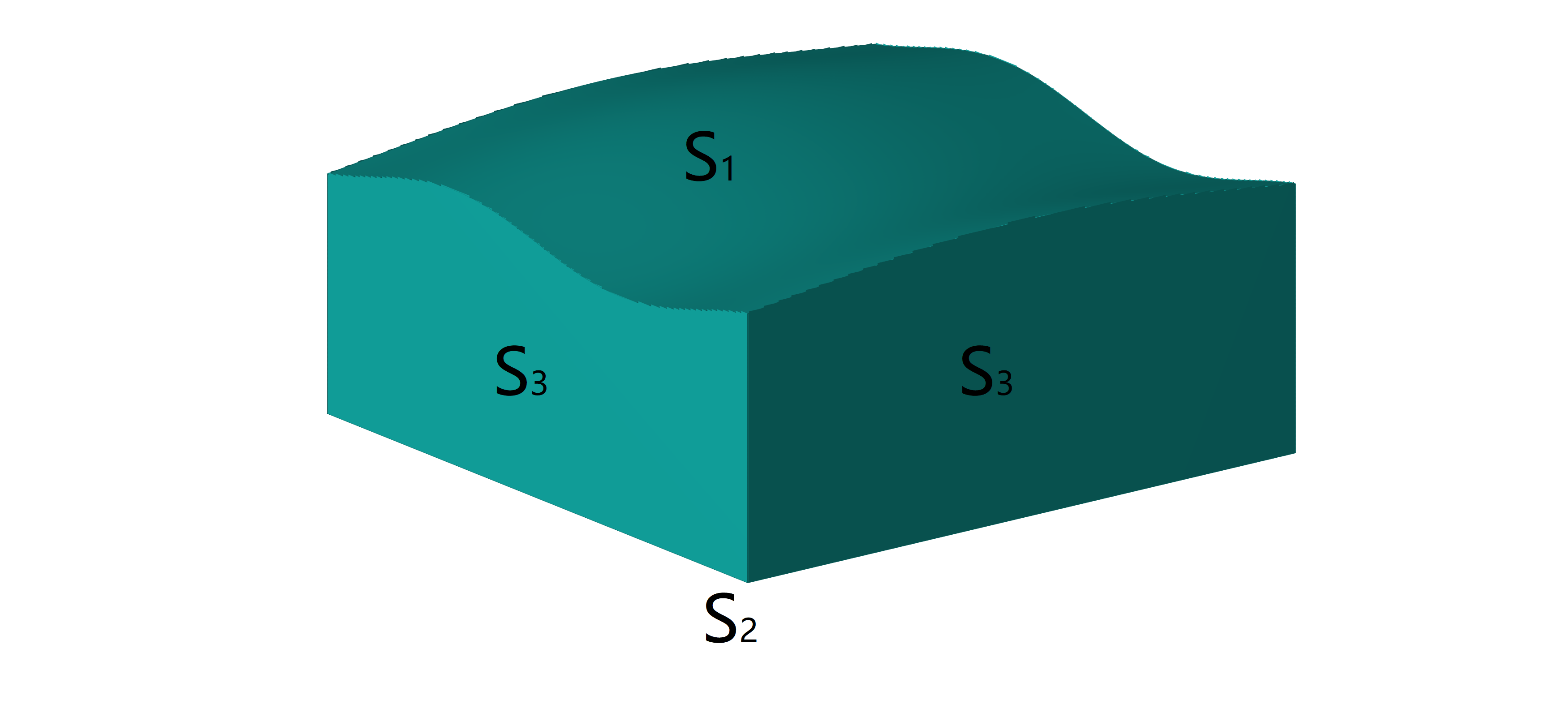}
\caption{A typical example of the 3D nonlinear heat conduction systems.}
\label{Fig2}
\end{figure}

The developed method aims to forecast a time-varying heat flux distribution of a nonlinear 3D heat conduction model with temperature-dependent heat conduction parameters. Fig. \ref{Fig2} displays the schematic of a typical example of such a model, where the solid body is a isotropic homogeneous domain denoted as $\Omega$. $S_1$ is a curved surface where the unknown temporal and spatial varying heat flux $q(r,t)$ is imposed. The bottom surface is designated as $S_2$ and the physical boundary is radiation boundary conditions. The rest of the surfaces are insulated and symbolized as $S_3$. The temperature sensors are uniformly scattered on the temperature surface $S_2$ which are denoted as $T_j\ (1\le j\le I)$, where $I$ is the total number of the sensors.

The governing equations, related boundaries and initial conditions of the system are

\begin{equation}
k\nabla^2T=\rho C_p\frac{\partial T}{\partial t},r\in\Omega,t\geq0
\end{equation}

\begin{equation}
-k\frac{\partial T(r;t)}{\partial z}=q(r;t),r\in {{S}_{1}},t\ge 0
\end{equation}

\begin{equation}
-k\frac{\partial T(r;t)}{\partial z}=\sigma \varepsilon \left( T_{w}^{4}(r;t)-T_{0}^{4} \right),r\in {{S}_{2}},t\ge 0
\end{equation}

\begin{equation}
-k\frac{\partial T(r;t)}{\partial n}=0,r\in {{S}_{3}},t\ge 0
\end{equation}

\begin{equation}
T(r;0)={{T}_{0}},r\in \Omega,t=0
\end{equation}
where $T(r;t)$ represents the temperature and $\rho $ designates the mass density of the solid body. Temperature-dependent heat capacity and thermal conductivity are designated as ${{C}_{p}}(T)$ and $k(T)$. And $\varepsilon $ is the emissivity while $\sigma $ is the Stefan Boltzmann Constant.

\begin{figure}[htbp]
\centering
\includegraphics[scale=.7]{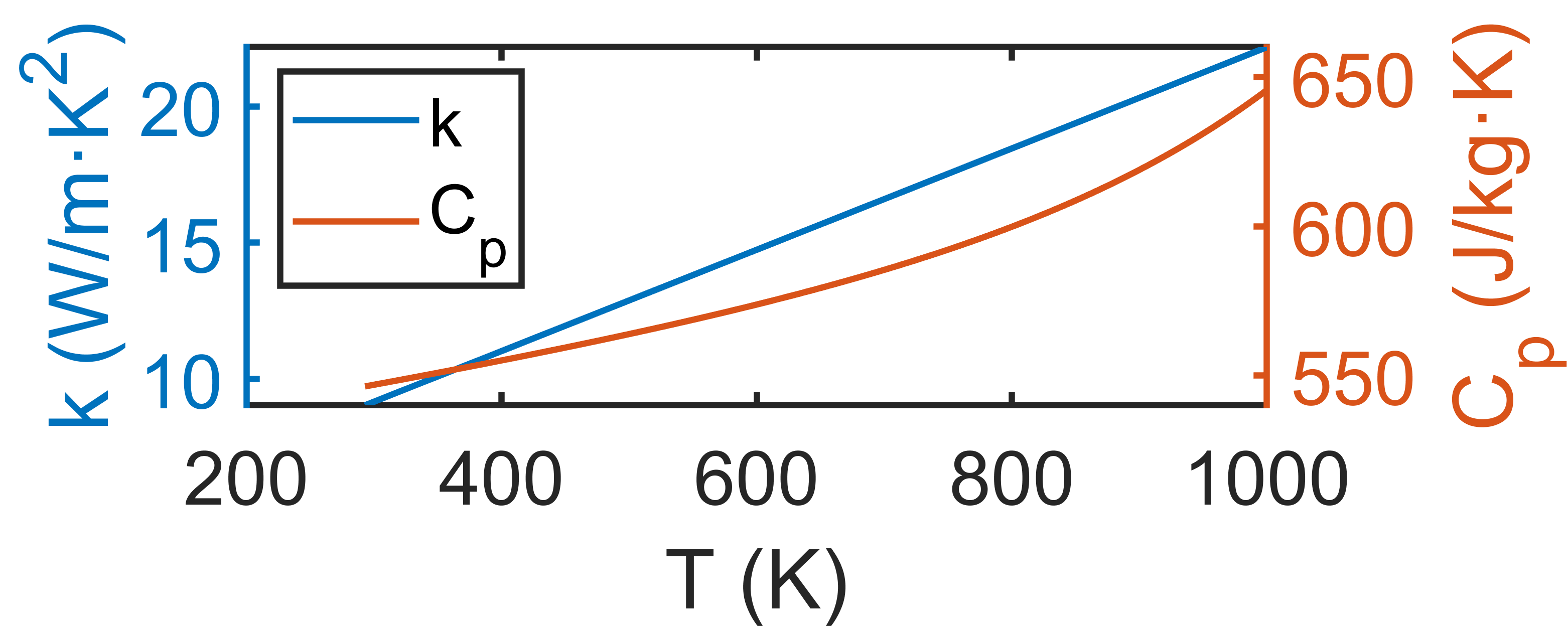}
\caption{The temperature-dependent thermal conductivity and heat capacity.}
\label{Fig3}
\end{figure}

The physic boundary settings, the detailed geometry parameters and thermal properties expressed in Tabel. \ref{Table1} will be the foundation of the test case. Fig. \ref{Fig3} shows the temperature-dependent thermal conductivity $k(T)$ and heat capacity ${{C}_{p}}(T)$ of an alloy used on the aircraft. Both the initial temperature of the solid body and ambient temperature are 273.15K. 900 temperature sensors are uniformly located on the bottom surface $S_2$.

\subsection{Data set generation}

After constructing the 3D nonlinear model, we evolve COMSOL based on the finite element method to generate sufficient data for training and testing. The detailed procedure of this algorithm is as follows:
Firstly, the 3D model is divided into $E$ sub grids, each of which contains $p$ nodes. The temperature at any given position $(x, y, z)$ can be expressed approximately by the interpolation of node temperature.

\begin{equation}
T = \sum\limits_{i = 1}^p {{N_i}(x,y,z){T_i}^e = \left[ N \right]{{\left\{ T \right\}}^{\left( e \right)}}}
\end{equation}
where $N_i$ is the weight of the $i_{th}$ node. Applying the functional theory, the heat conduction equation can be rewritten as
\begin{equation}
\left[ {{K_1}^{(e)}} \right]{\left\{ T \right\}^{\left( e \right)}} + \left[ {{K_2}^{(e)}} \right]{\left\{ T \right\}^{\left( e \right)}} + \left[ {{K_3}^{(e)}} \right]{\left\{ {\dot T} \right\}^{\left( e \right)}} - {\left\{ P \right\}^{\left( e \right)}} = 0
\end{equation}
where $\left[ {{K_1}^{(e)}} \right]$, $\left[ {{K_2}^{(e)}} \right]$, $\left[ {{K_3}^{(e)}} \right]$ and $\left\{ P \right\}^{\left( e \right)}$ denote the heat conduction, the boundary condition, the unsteady and the load matrix, respectively. By assembling each sub matrix, it can be obtained that
\begin{equation}
\left[ K \right]\left\{ {\overline T } \right\} = \left\{ {\overline P } \right\}
\end{equation}
where the coefficient matrix are
\begin{equation}
\left[ K \right] = \sum\limits_{e = 1}^E{\left( {\left[ {{K_1}^{\left( e \right)}} \right] + \left[ {{K_2}^{\left( e \right)}} \right]} + \left[ {{K_3}^{\left( e \right)}} \right]\right)}
\end{equation}

\begin{equation}
{\left\{\overline P \right\}}  = \sum\limits_{e = 1}^E {{{\left\{ P \right\}}^{\left( e \right)}}}
\end{equation}

After imposing (2)-(5) to (8), the temperature at the $S_2$ can be acquired. With various models generated by MATLAB, it is feasible to produce sufficient data for training and testing.

\subsection{Architecture of the framework}
The 3D mode in our model has two tough points: (1) there are 900 sampling points of heat flux to be predicted, so that it is inefficient to use fully connected-LSTM to predict one by one; (2) there are various shapes of the solid body, so that the temperature distribution is influenced by both the heat flux and the body shapes. To address these problems, the ConvLSTM \cite{shi2015convolutional}, containing both convolutional structures in input-to-state and state-to-state transitions, is introduced to the framework. By assembling multiple ConvLSTM layers with LSTM and fully connected layers, we could build a network model for 3D IHCPs.

A major drawback of fully connected-LSTM in handling spatiotemporal data is that it uses only fully connected layers in transitions so that it does not encode any spatial information. However, since all the inputs ${{\chi }_{1}},\cdots ,{{\chi }_{t}}$, cell outputs ${{C}_{1}},\cdots ,{{C}_{t}}$, hidden states ${{H}_{1}},\cdots ,{{H}_{t}}$ and gates ${{i}_{t}},{{f}_{t}},{{o}_{t}}$ of the ConvLSTM are 3D tensors , it is feasible to store different spatial information in different channels. This is achieved by using a convolution operator in the transitions. The key equations of ConvLSTM are shown below, where `$*$' denotes the convolution operator and `$\circ$' denotes the Hadamard product:

\begin{equation}
{{i}_{t}}=\sigma ({{W}_{xi}}*{{\chi}_{t}}+{{W}_{hi}}*{{H}_{t-1}}+{{W}_{ei}}\circ {{C}_{t-1}}+{{b}_{i}})
\end{equation}

\begin{equation}
{{f}_{t}}=\sigma ({{W}_{xf}}*{{x}_{t}}+{{W}_{hi}}*{{H}_{t-1}}+{{W}_{cf}}\circ {{c}_{t-1}}+{{b}_{f}})
\end{equation}

\begin{equation}
{{C}_{t}}={{f}_{t}}\circ {{C}_{t-1}}+{{i}_{t}}\circ \tanh ({{W}_{xc}}*{{x}_{t}}+{{W}_{hc}}*{{h}_{t-1}}+{{b}_{c}})
\end{equation}

\begin{equation}
{{o}_{t}}=\sigma ({{W}_{xo}}*{{x}_{t}}+{{W}_{ho}}*{{h}_{t-1}}+{{W}_{co}}\circ {{c}_{t}}+{{b}_{o}})
\end{equation}

\begin{equation}
{{H}_{t}}={{o}_{t}}\circ \tanh ({{C}_{t}})
\end{equation}

\begin{figure}[htbp]
\centering
\includegraphics[scale=1.3]{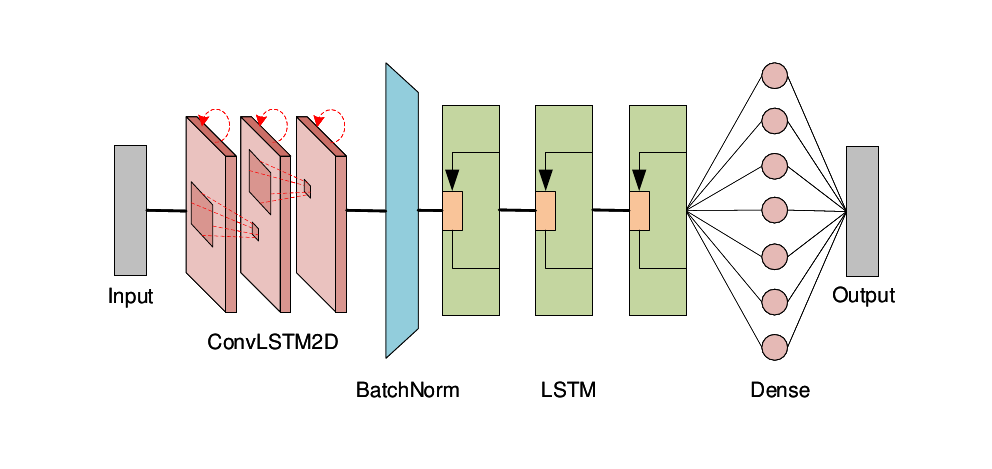}
\caption{Architecture of the framework in solving inverse heat conduction problems.}
\label{Fig4}
\end{figure}

ConvLSTM can serve as a layer for a more complex framework to capture spatial-temporal characteristics in practical scenarios. In this work, the proposed framework shown in Fig. \ref{Fig4} with three layers of ConvLSTM, three layers of FC-LSTM and one layer of fully connected network is applied to solve the complex IHCPs. By minimizing the kernel size gradually in different layers, this structure could capture the different sizes of spatial relationships efficiently. We set two kernel filters of the first layers as $4\times 4$ to separately keep the geometry shapes and temperature distributions. Considering the nonlinearity of the system, we add an extra layer in which the kernel filter size is $2\times 2$ to better fit the spatial influence of the heat flux. And the last layer of ConvLSTM is to unify the result of two filters into one. By minimizing the kernel filters and unifying two channels, the output of the last ConvLSTM layer is a combination of geometry shape and temperature. Therefore, the input of LSTM could be imagined as a 3D temperature which is similar to the heat flux in spatial so that we only need to capture the temporal information. The first two layers of LSTM are set to return the sequence in each time step, as this will help find the relationship between the temperature and heat flux more precisely. At the last layer, the LSTM is used to set the sensitive coefficient for every time step to make sure the main heat flux influence the temperature at this time. To avoid missing the information included in former layers, the number of the neurons is the same as that of the sampling points of heat flux. After the LSTM, we also add a  fully connected layer to resize the output of the LSTM as it may be unable to match the supervised data. Since the network has multiple stacked ConvLSTM layers, it has striking power to give predictions in complex dynamic systems like IHCP in 3D complex models.

\subsection{Training process}

\begin{figure}[htbp]
\centering
\includegraphics[scale=0.7]{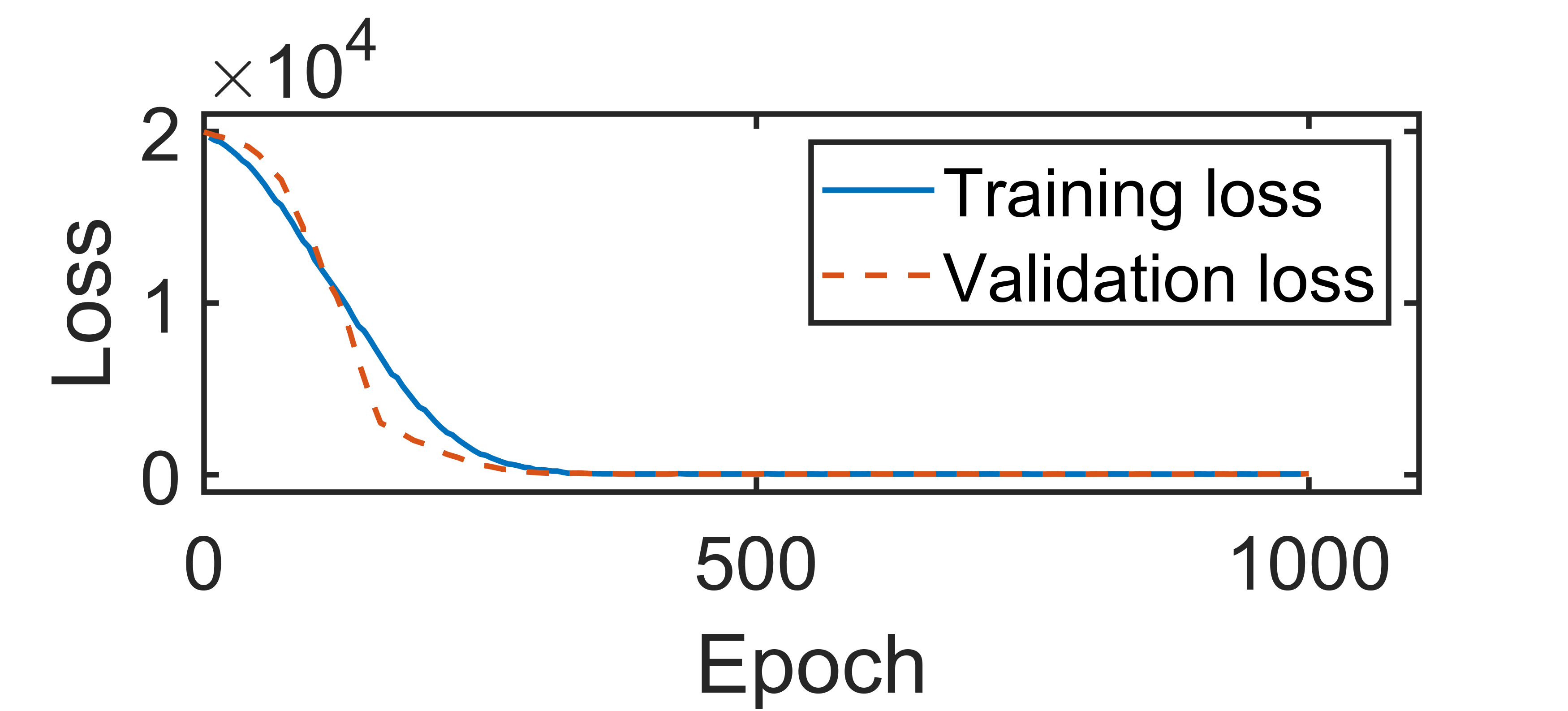}
\caption{The loss curve of the training process.}
\label{Fig5}
\end{figure}

The training process is performed on an NVIDIA RTX 2080 Ti graphic card on Dell Precision 7920 Tower. During the process, the loss of the training set (training loss) and that of the test set (validation loss) decrease gradually. After 300 iterations, the loss reduced to an acceptable level. The loss curves are exhibited in Fig. \ref{Fig5}.

\section{Results}
In this part, some specimens with various heat flux distributions are chosen from the complex 3D data set to assess the performance of the proposed framework. Here, the ground truth of heat flux distribution is defined by the data generated through MATLAB. 

\subsection{Regular 3D models}

\begin{figure}[htbp]
\centering
\includegraphics[scale=.4]{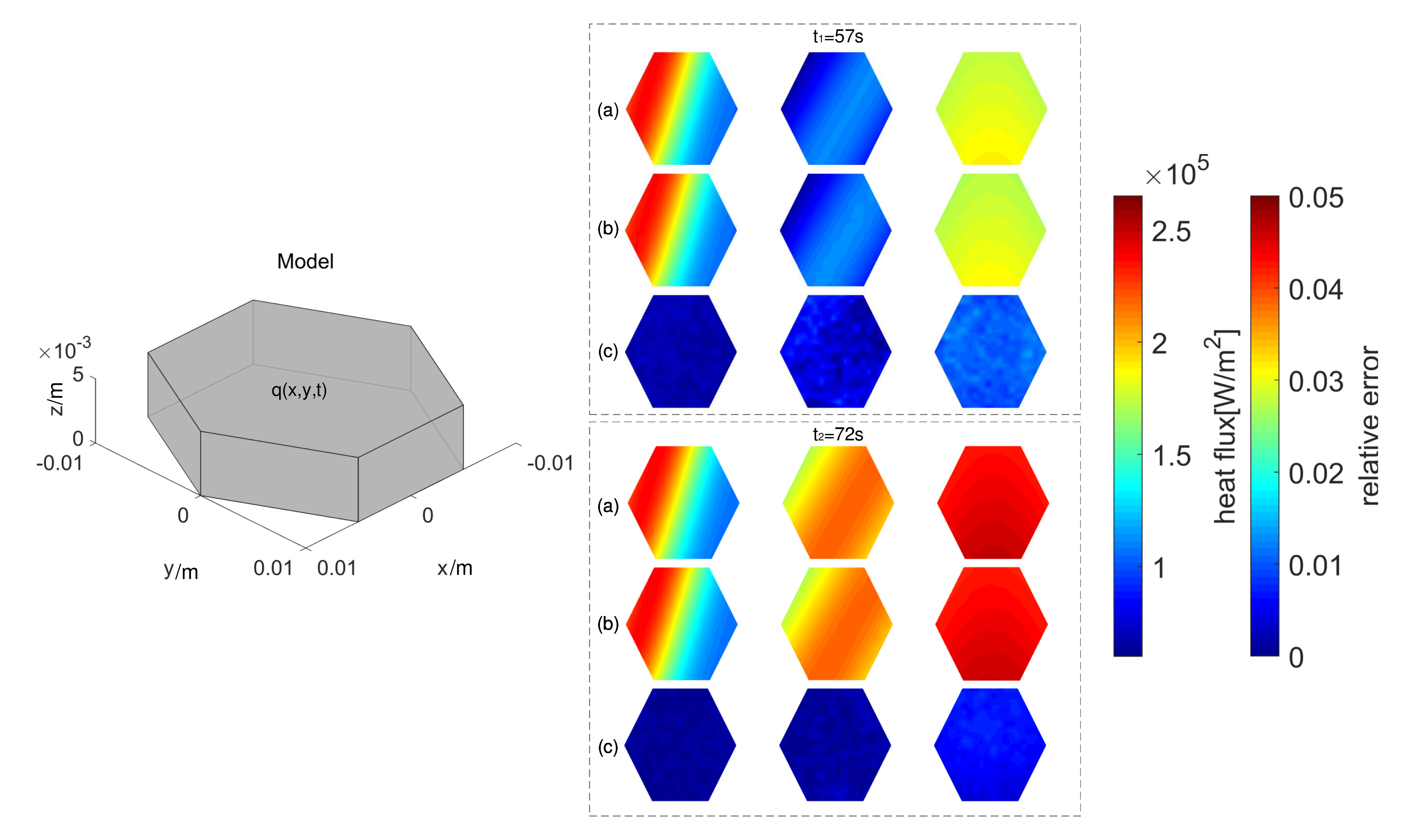}
\caption{Samples in the regular 3D test set. (a) The predicted heat flux, (b) the ground truth of heat flux and (c) the relative errors.}
\label{Fig6}
\end{figure}

Before starting to handle 3D complex models, we first build up a simple 3D model with flat surfaces to evaluate the ability of the proposed framework in capturing the temporal and spatial features. In such models, as the heat flux surface and the temperature measured surface are parallel, the spatial distribution of the temperature is similar to that of the heat flux and this kind of models is denoted as regular. The main difficulty is capturing the temporal features. Here we use three heat flux cases to test the network and the outcome is visualized in Fig. \ref{Fig6}. We randomly choose two different time steps to assess the framework and the result exhibits that the difference in each model is slightly affected by the time steps which demonstrates the proposed framework is good at predicting the temporal variation of the heat flux. To show the precise difference between the ground truth of heat flux and predicted ones, we add the distribution of relative error in each case which is defined as

\begin{equation}
error=\left| \frac{{{q}_{predicted}}-{{q}_{real}}}{{{q}_{real}}} \right|
\end{equation}

\subsection{Complex 3D models}

After the 3D regular models, we change the training data of the framework and measure the ability of the ConvLSTM to reconstruct the distribution of heat flux combining the features of complex 3D models and temperature distributions. In this kind of models which is symbolized as complex, the heat flux surface is no longer parallel to the temperature measured surface and contains curved surface. To demonstrate the performance of the ConvLSTM framework in a better way, we randomly choose twelve cases to exhibit the temporal and spatial inversion ability separately. Fig.6 shows three cases that include different complex models and heat flux distributions. The geometry shapes of the models varies in many different aspects. The max height, min height and curvature are different while the relative error is stable. Therefore, we could figure out that the proposed  framework performed pretty well on complex 3D models and have a good generalization ability in geometry shapes.

\begin{figure*}[htbp]
\centering
\includegraphics[scale=.221]{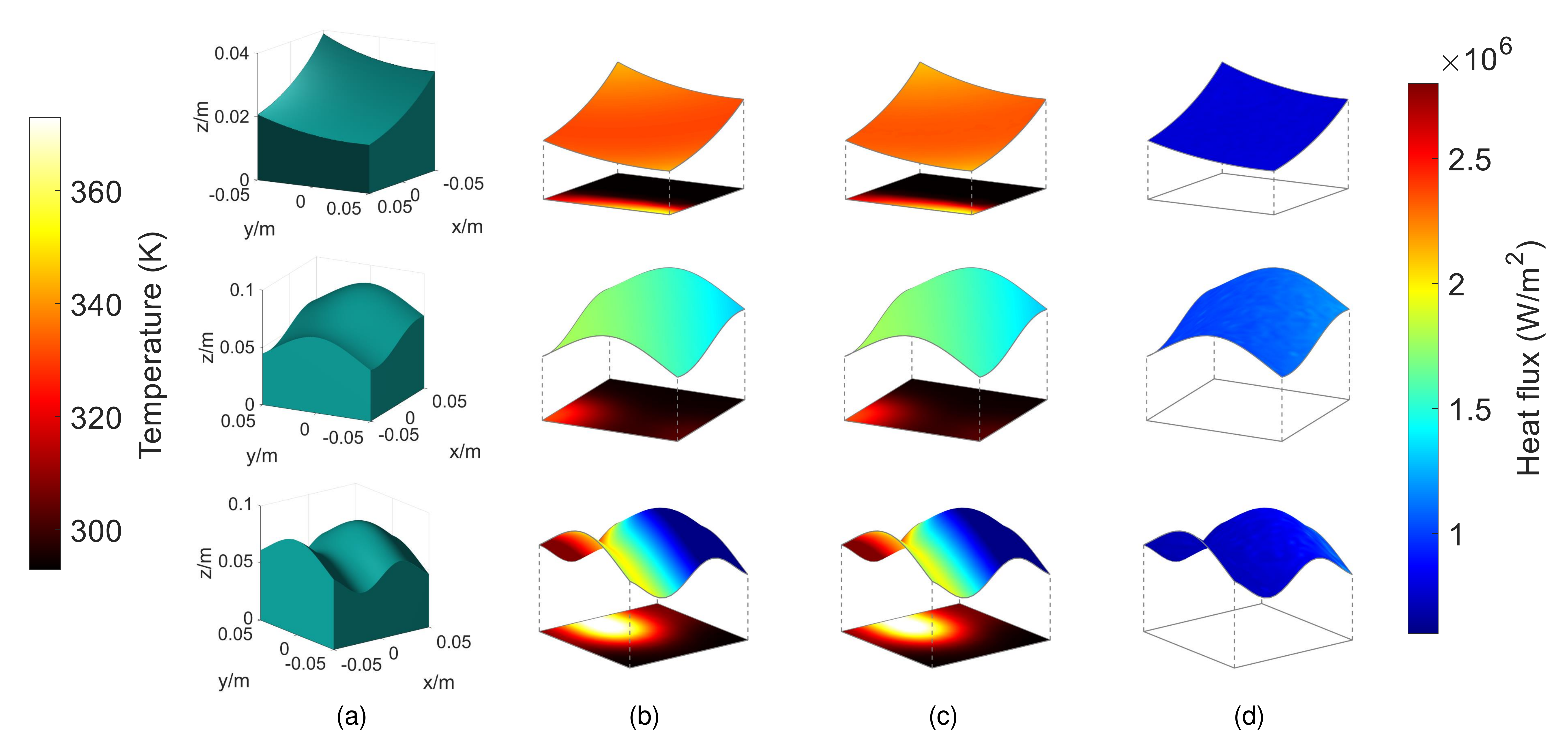}
\caption{Samples in the complex 3D test set. (a) The geometry of the complex models, (b) the forecast of the heat flux and the measured temperature (t=70s) (c) the ground truth of the heat flux and the measured temperature (t=70s) and (d) the relative errors.}
\label{Fig7}
\end{figure*}

\begin{figure*}[htbp]
\centering
\includegraphics[scale=.22]{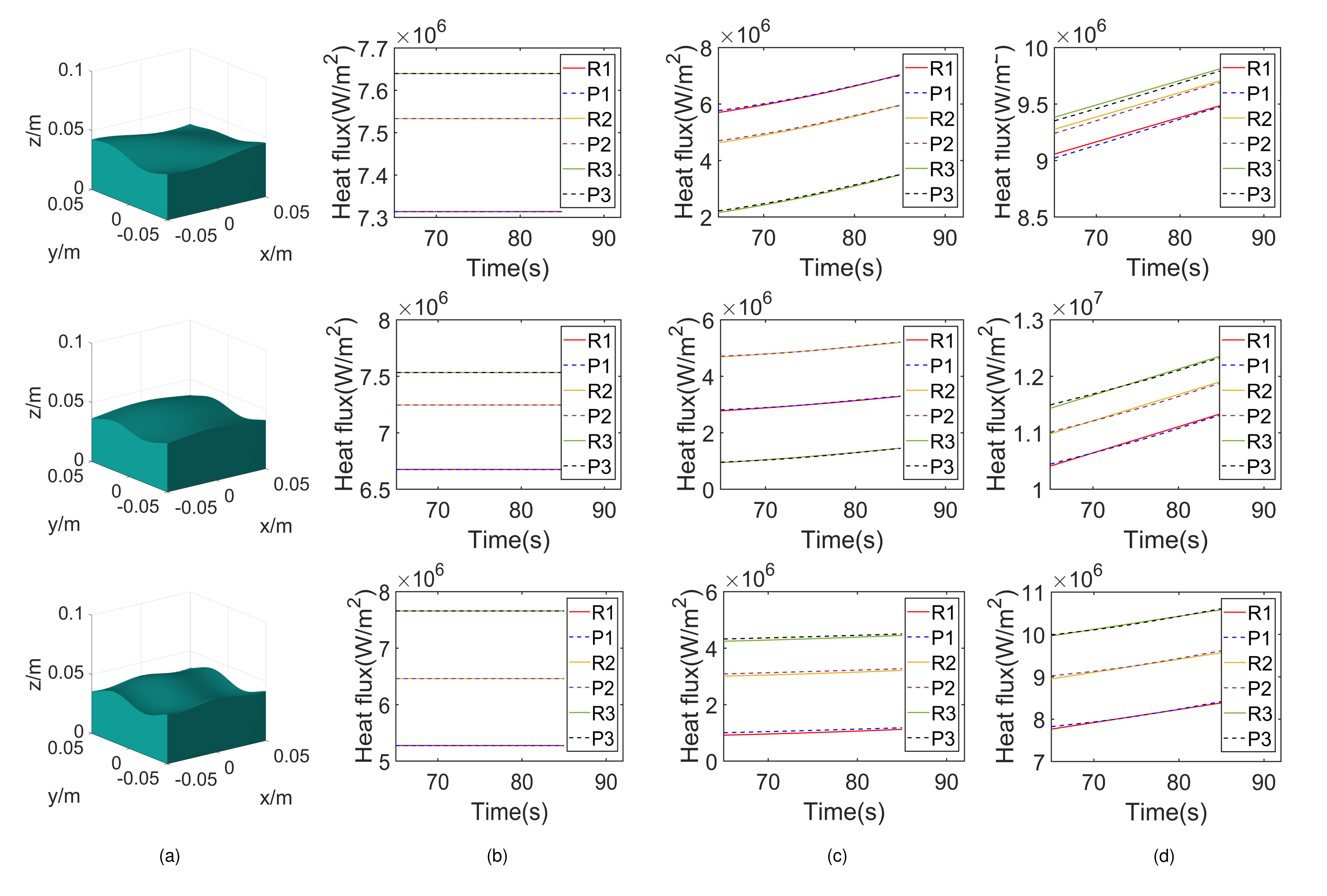}
\caption{Samples in the complex 3D test set. (a) The geometry of the complex models, (b), (c) and (d) three different heat flux distributions imposed on (a). $R_i (i=1, 2, 3)$: ground truth, $P_i (i=1,2,3)$: prediction.}
\label{Fig8}
\end{figure*}

In Fig. \ref{Fig8}, there are three complex models and three different heat fluxes for each model. These heat fluxes are unevenly distributed in space, and we randomly select three points to illustrate the temporal fitting capability of the framework. The three points are located on the diagonal of the surface and have different ranges of the heat flux variation. The result shows that the range of the heat flux variation and location of the point have few influence on the prediction and demonstrates that our framework has a good generalization ability in predicting temporal and spatial related heat flux.

\subsection{Calculation accuracy and speed}

\begin{figure}[htbp]
\centering
\includegraphics[scale=.5]{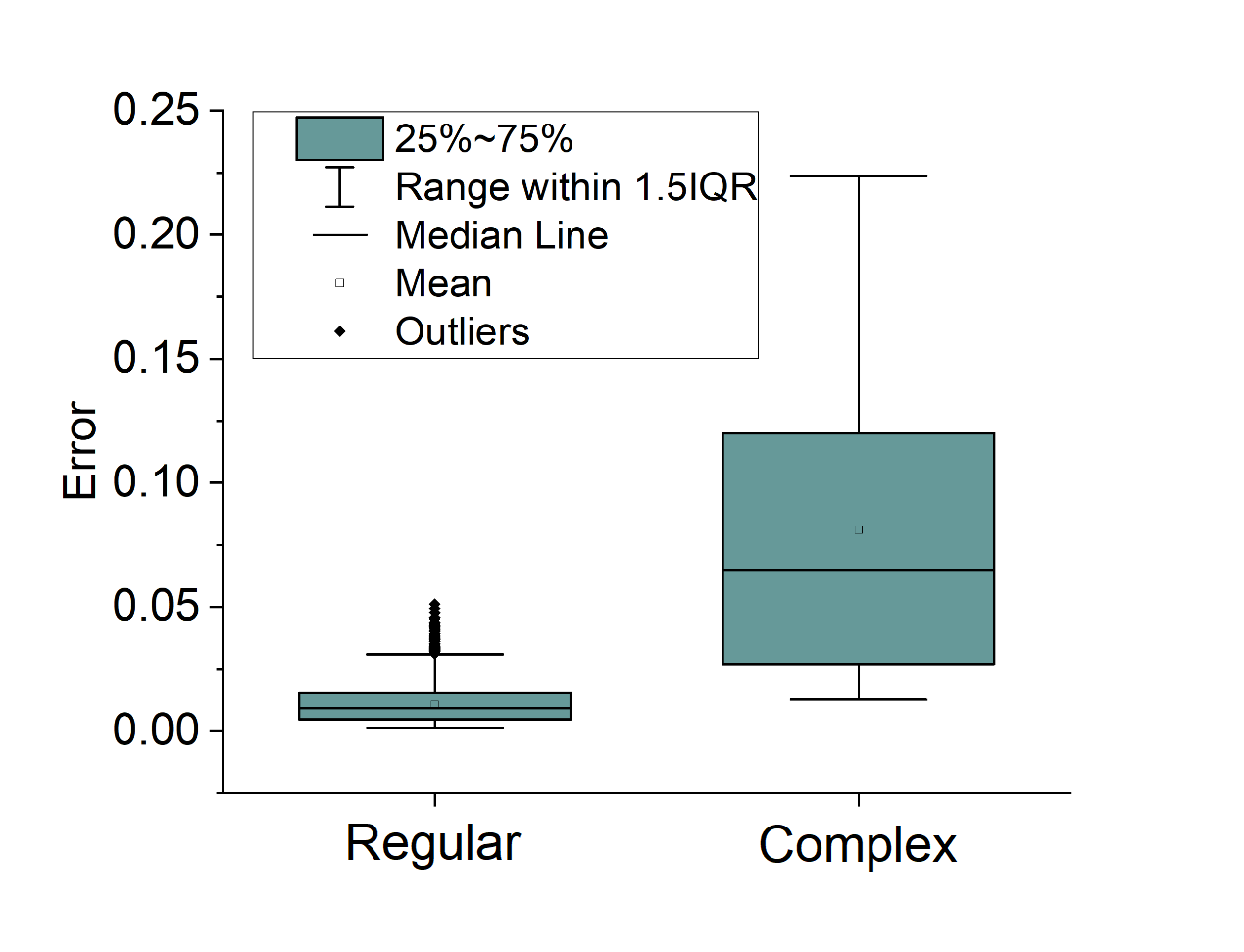}
\caption{Statistical distribution for test cases separately produced by two kinds of generators. The regular one represents the flat heat flux surface while the complex one represents the complex surface.}
\label{Fig9}
\end{figure}

We compute the relative error for the test case in Fig. \ref{Fig9}. As mentioned before, the regular refers to the regular models of which the two surfaces are parallel while the complex denote the complex models where two surfaces are not parallel. Since the regular models offer more information about the heat flux, the difficulty of the IHCP for regular ones is less and the loss distribution is better than the complex ones. These figures illustrate that for regular 3D models it is easier to predict the heat flux more precisely and stable.

Also, since the framework is based on the GPU and CUDA is employed to forecast the heat flux, the average time cost to predict a case which contains 71 time steps and represents a 35s period is only 1.27s. In other words, for each time step, it takes time within 20 ms.

\section{Conclusion}

In this work, we have proposed a new framework to forecast the heat flux of complex 3D models, which have temperature-dependent thermal parameters and nonlinear physical boundaries. We introduce a new variation of ConvLSTM which is designed to solve spatial and temporal-related problems. Compared to the regular iterative methods, such a framework can work efficiently after being trained properly. As it usually uses CUDA to calculate in parallel, it could figure out the heat flux preciously within 20 ms, which is extremely superior to those traditional algorithms. Also, as this framework combines ConvLSTM with LSTM, it takes few epochs to reach convergence. The result illustrate that our framework have a good generalization ability in forecasting temporal and spatial related heat flux. Hence, the proposed framework can replace the direct heat flux measurements in real-time heat flux inversion problems with advanced temperature measurements and have multiple applications in many industrial scenarios like engine design or furnace operation in the future.

\section*{Acknowledgments}
This work is supported in part by the National Nature Science Foundation (Grant No. 61801009), in part by the National Science and Technology Major Project (Grant No. 2019-VIII-0009-0170), in part by Defense Industrial Technology Development Program (Grant No. JCKY2020601B011) and in part by Stable Operation Project.

\bibliographystyle{unsrt}  

\bibliography{ref.bib}

\end{document}